\documentclass[12pt]{article}
\usepackage{amsmath}
\usepackage{amsfonts}
\usepackage{amssymb}
\usepackage{amsthm}
\usepackage{indentfirst}
\usepackage{fancyhdr}
\usepackage{graphicx}
\usepackage{float}

\usepackage[ruled, linesnumbered]{algorithm}
\usepackage{algorithmicx}
\usepackage[noend]{algpseudocode}

\usepackage[title]{appendix}

\usepackage[utf8]{inputenc}

\usepackage{multirow}

\usepackage[
backend=biber,
style=alphabetic,
sorting=ynt
]{biblatex}

\newtheorem{theorem}{Theorem}[section]

\newtheorem{lemma}{Lemma}[section]

\usepackage[textsize=tiny]{todonotes}

\title{Simulated Quantum Annealing is Efficient on the Spike Hamiltonian}
\author{Thiago Bergamaschi}
\date{April 2020}

\begin{document}
\maketitle

\textbf{Abstract}\\

In this work we study the convergence of a classical algorithm called Simulated Quantum Annealing (SQA) on the Spike Hamiltonian, a specific toy model Hamiltonian for quantum-mechanical tunneling introduced by \cite{Farhi2002QuantumAE}. This toy model Hamiltonian encodes a simple bit-symmetric cost function $f$ in the computational basis, and is used to emulate local minima in more complex optimization problems. In previous work \cite{Crosson2016SimulatedQA} showed that SQA runs in polynomial time in much of the regime of spikes that QA does, pointing to evidence against an exponential speedup through tunneling. In this paper we extend their analysis to the remaining polynomial regime of energy gaps of the spike Hamiltonian, to show that indeed QA presents no exponential speedup with respect to SQA on this family of toy models.

\section{Introduction}
Ever since the seminal work of \cite{Farhi2000QuantumCB}, Quantum Annealing, or Quantum Adiabatic Optimization at low temperatures, has proved to be powerful heuristic optimization algorithm. The key intuition behind this quantum algorithm is to encode the minimizer of some objective function $f:\{0, 1\}^n\rightarrow \mathbb{R}$ over $n$ bit bitstrings into the ground state of a diagonal ``cost" Hamiltonian $H_f = \sum_{x\in\{0, 1\}^n}f(x)|x\rangle \langle x|$, and then to slowly relax the system into the ground state we are interested in by interpolating the system Hamiltonian with an ``easy to prepare" initial Hamiltonian. In contrast to the rich theoretical literature on this annealing process and its algorithmic applications, practical limitations on quantum hardware imply that very little is known on the empirical performance of these algorithms. This motivates the need for formal, provable correctness results and theoretical guarantees, and in particular verification of any reported exponential algorithmic speedups.

\cite{Farhi2002QuantumAE} and \cite{Reichardt2004TheQA} were the first to compare quantum annealing (QA) with its direct classical analog, simulated annealing (SA), and moreover showed an exponential separation between the runtimes of these two procedures on specific toy model systems. The key physical insight from their results is that the quantum-mechanical wave-function would be able to exploit symmetries, or tunnel through certain barriers in the energy landscape, whereas the thermal process would be stuck in a local minima, unable to jump over the barrier.

One of the reported exponential speedups by \cite{Farhi2002QuantumAE}, and the one we will be focusing on in this paper, is the \textit{Spike Hamiltonian}. This toy model Hamiltonian encodes a very simple local minima, and emulates quantum-mechanical tunneling through a barrier of polynomial height and polynomial width in the energy landscape. In particular, the objective function is

\begin{equation}
    f(w) = \begin{cases}
       |w|+n^\alpha & \text{if } w\in R \\
       |w| & \text{otherwise}
    \end{cases}
\end{equation}
where the Spike Region $R$ is defined in terms of the hamming weight of the input $w$
\begin{equation}
    R = \bigg\{w\in \{0, 1\}^n \text{ s.t. } \frac{n}{4}-\frac{n^\eta}{2}\leq |w|\leq \frac{n}{4}+\frac{n^\eta}{2}\bigg\}
\end{equation}
which effectively says that if the hamming weight $|w|$ of the bitstring input $w\in \{0, 1\}^n$ is inside the a region of polynomial width $n^\eta$ around $n/4$, then the energy ``spikes" by $n^\alpha$. An unpublished, folklore result by Goldstone, later formally verified by \cite{Brady2016SpectralgapAF}, pointed out three very different types of Spike Hamiltonians, parametrized in terms of the constants $0\leq \alpha, \eta < 1$. If we denote as $\Delta$ as the minimum energy gap, the difference between the first excited state and the ground state of the quantum system during the execution of the quantum annealing algorithm, then in these three regions $\Delta$ has very different asymptotic dependencies, parametrized in terms of very simple linear inequalities over $(\alpha,\eta)$:
\begin{equation}
    \Delta \propto \begin{cases}
       \text{constant} & \text{if } \alpha+\eta \leq 1/2 \\
       n^{1/2-\alpha-\eta} & \text{if } \alpha+2\eta \leq 1 \text{ and } \alpha+\eta > 1/2 \\
       \text{poly}(n)\exp\{-\Omega(n^{\alpha+2\eta-1})\} & \text{otherwise}
    \end{cases}
\end{equation}
which we will refer to as the constant, polynomial and exponential regimes. While Quantum Annealing converges in polynomial time in the polynomial regime, and even in linear time in the constant regime \cite{Reichardt2004TheQA}, Simulated Annealing takes exponential time to sample from the ground state in every regime above.

The focus of this work is a classical algorithm inspired by both SA and QA. Quantum Monte Carlo (QMC) methods are classical, Markov Chain Monte Carlo (MCMC) algorithms that exploit the quantum-to-classical mapping of Suzuki et al \cite{Suzuki1977MonteCS} \cite{Suzuki1986QuantumSM}. When QMC methods are applied to simulating the evolution of low-temperature Gibbs states of quantum annealing (QA) Hamiltonians, the resulting algorithm is called Simulated Quantum Annealing (SQA). Much like SA, and other applications of MCMC algorithms to Simulations, Statistics, and Optimization, making provable theoretical convergence guarantees on SQA is a difficult task. 

Comparing the computational power of QA and SQA is still a largely open question. Quantum Adiabatic Optimization, as defined by slowly interpolating the system between two arbitrarily Hamiltonians is known to be BQP complete \cite{Aharonov2004AdiabaticQC}, however when restricted to Hamiltonians with the ``no sign problem" seems to loose a lot of its computational power. The standard formulation of QA is based on Hamiltonians with the no sign problem, or Stoquastic Hamiltonians, i.e. those with strictly non-positive off-diagonal entries. \cite{Bravyi2006MerlinArthurGA}, \cite{Bravyi2008TheCO} studied the Local Hamiltonian problem to show that QA under stochastic Hamiltonians is in PostBPP, pointing to evidence of a separation from BQP. \cite{hastings2013obstructions} constructed multiple examples where QA converges in polynomial time where SQA takes exponential time, and later in \cite{hastings2020power} showed a super-polynomial oracle separation between the power of adiabatic quantum computation with no sign problem (and a polynomial energy gap) and the power of classical computation.

Recently, \cite{Crosson2016SimulatedQA} showed a constructive upper bound for the mixing time of an SQA algorithm on the Spike Hamiltonian, proving that SQA converged in polynomial time over the constant energy gap region. This pointed out that the classical simulation of the quantum process was actually exponentially faster than the classical, thermal, SA process and additionally presented theoretical evidence for a correlation between QA and SQA results that had been previously experimentally observed on D-Wave's actual quantum hardware \cite{Boixo2014EvidenceFQ}. Later, the work of \cite{Jiang2017ScalingAA} used very different techniques on the exponential energy gap regime, but came to a similar conclusion of the equivalence between the QMC escape rate and QA tunneling rate through the barrier. 

Our intention in this paper is to extend the work of \cite{Crosson2016SimulatedQA} to the polynomial gap regime, by modifying their analysis of the convergence of SQA to show that it indeed allows us to sample from the ground state of the Spike Hamiltonian with a polynomial energy gap in polynomial time. In this manner, we show that on this toy model for quantum mechanical tunneling SQA inherits many of the algorithmic advantages of quantum dynamics, and in particular there is no exponential speedup for sampling from the ground state of the Spike Hamiltonian. Essentially, the standard version of SQA that does not use any structure of the problem finds the minimum of the spike cost function as fast as QA. Formally,

\begin{theorem} \label{mainresult}
Simulated Quantum Annealing based off of the path-integral Monte Carlo method efficiently samples the output distribution of QA for the spike cost function when $\alpha+2\eta\leq 1$. The running time using single-site spin flips is $O(n^{37})$.
\end{theorem}

\subsection{Previous Work}
A number of previous papers have analysed numerically and analytically the spike Hamiltonian in different regimes of spectral gaps, however, mostly in the context of comparing QA and SA. \cite{Farhi2002QuantumAE} and \cite{Reichardt2004TheQA} were the first to show an exponential separation between QA and SA, however limited to the region of constant energy gap, $\alpha+\eta\leq 1/2$. More recently, \cite{Kong2015ThePO} and \cite{Brady2016SpectralgapAF} formalized Goldstone's folklore result and showed precise asymptotic expressions for the energy gap in the other relevant regions of $(\alpha,\eta)$ parametrizations.

\cite{Crosson2014TunnelingTH}, \cite{Crosson2016SimulatedQA}, \cite{Jiang2017ScalingAA}, \cite{Crosson2018RapidMO} were the first to numerically or analytically compare Quantum Monte Carlo Methods with Quantum Annealing. \cite{Jiang2017ScalingAA} used instanton methods to analytically show that the exponential scaling of the thermally-assisted tunneling rate and the escape rate of the QMC process on a fully connected quantum spin model are the same. Their results imply exponentially large lower bounds on the runtime of the ``large barrier" $\alpha+2\eta>1$ exponential regime of spike hamiltonians, and additionally, point out that on the interface of the energy gap regimes where $\alpha+2\eta = 1$ and $\alpha \geq \eta$ (height larger than width) both QA and SQA have a polynomial lower bound to their ``tunneling" rates.

\cite{Crosson2016SimulatedQA} showed a constructive polynomial upper bound on the mixing time in the constant gap regime, where $\alpha+\eta\leq 1/2$, by using the PIMC method to develop a Markov Chain that samples in polynomial time from a distribution polynomially close in variational distance from the ground state of the Spike Hamiltonian. Our result in this paper is to ``fill in the gap" of energy gaps by extending the analysis of  \cite{Crosson2016SimulatedQA} to show a polynomial time algorithm for the entirety of the polynomial energy gap regime of $\alpha+2\eta\leq 1$.

\subsection{Technical Overview and Our Contribution}
Our intention is to construct a classical MCMC algorithm that simulates thermal QA and allows us to sample from the ground state, so we begin by bounding the total variational distance between the ground state and the low temperature Gibbs state $\sigma\propto e^{-\beta H}$ of the actual quantum procedure. We use the very loose upper bound of Lemma \ref{thermalerror} to show that in the polynomial gap regime, it suffices to pick an inverse temperature that is polynomially colder $\beta = \Omega(n^{\frac{1}{2}+\alpha+\eta})$ than that used in the approach of \cite{Crosson2016SimulatedQA} to establish the same exponentially decreasing distance guarantees, that is, 

\begin{equation}
 ||\sigma-|\psi_0\rangle\langle\psi_0|||_1\leq e^{-\Omega(n)}
\end{equation}

Once the temperature has been established, in Sections 2.3 and 2.4 we follow the approach of \cite{Crosson2016SimulatedQA} to define the \textit{single-site Metropolis updates} Markov Chain whose stationary distribution is approximately the thermal distribution $\Pi(z) = |\langle z| \sigma|z\rangle|^2$ we wish to sample from, via the quantum-to-classical mapping of Suzuki et al \cite{Suzuki1977MonteCS}, \cite{Suzuki1986QuantumSM}. This reduces the analysis to showing the efficient convergence of the underlying Markov Chain, and in Section 3 we point out the relevant machinery and details of the Most-Paths Comparison method originally developed by \cite{Crosson2016SimulatedQA} for the constant gap regime. 

In particular, the approach hinges on the idea of treating the spike as a perturbation, and comparing the spike and the spike-less Markov Chains $(\pi, P)$ and $(\tilde{\pi}, \tilde{P})$ respectively. These chains are defined over the same state space $\Omega = \{0, 1\}^{\text{poly}(n)}$ and adjacency graph $E$, however, comparing them faces the fundamental issue of an exponentially large number of states $x\in \Omega$ that are exponentially less likely in one of the chains over the other - i.e $\pi(x)/\tilde{\pi}(x)\ll 1$. This is addressed by partitioning the state space based on that stationary probability ratio
\begin{equation}
    \Omega_\theta =\bigg\{ x\in \Omega \text{ s.t. }\frac{\pi(x)}{\tilde{\pi}(x)}<\theta\bigg\}
\end{equation}

\noindent for some later determined $\theta = O(1)$, and constructing a second partition $\Omega_G\cup \Omega_B=\Omega$ of the state space into ``good" and ``bad" states, where the ``good" states are well connected in both chains, and they concentrate the stationary distribution in the spike chain. The key conclusion of the machinery of \cite{Crosson2016SimulatedQA} for our purposes is that the TV distance between the random walk and the stationary distribution is the composition of an exponential decay due to a fast mixing over the ``good" states, and a constant rate leakage to the ``bad" states, such that
\begin{equation} \label{mixing}
||\mu P^t-\pi||_1\leq Mt\pi(\Omega_B) + \pi_{min}^{-1}e^{-t/\rho}
\end{equation}

which is formally detailed in Theorems \ref{mostpaths} and \ref{coupling} in the discussion in Section 3. Here, the constant $M$ is defined by the warm start initialization, and $\rho = O(\theta \tilde{\rho})$ is the (polynomially large) congestion over the good subset in the spike chain, but we defer these discussions to Sections 3 and 5 as they don't essentially differ from \cite{Crosson2016SimulatedQA}'s presentation. Returning to equation \ref{mixing}, what this implies is that so long as the leakage probability $\pi(\Omega_B)$ is sufficiently small, there is an intermediary \textit{quasi-stationary} time period where the total variation distance above is actually exponentially small, thereby allowing us to sample from the stationary distribution with negligible error.

The main technical contribution in this paper is to tighten the analysis of the leakage probability to show that indeed there is a certain parametrization $\theta$ s.t. the leakage is exponentially small, which we later formalize in Lemma \ref{leakageprob} and present here:

\begin{lemma}[The Leakage Probability]
When $\alpha+2\eta \leq 1$ there exists a threshold $\theta = \Theta(1)$ such that the leakage probability in the spike-less is negligible, that is, $\tilde{\pi}(\Omega_\theta)\leq e^{-\Omega(n)}$.
\end{lemma}

We do so by revisiting the original concentration bound of \cite{Crosson2016SimulatedQA}. The key insight there was that a given state $x \in \{0, 1\}^{nL}$ is \textit{not} in $\Omega_\theta$ if 
\begin{equation}
   x\notin \Omega_\theta \iff \frac{\pi(x)}{\tilde{\pi}(x)} \geq \theta \iff ST(x)\equiv \sum_i \mathbb{I}_{x_i\in S}\leq b_\theta \equiv O(\frac{L}{\beta n^\alpha}\ln \theta^{-1})
\end{equation}
where $ST(x)$ is the \textit{spike time} of the state $x$, i.e. if the number of time-slices $x_i\in \{0, 1\}^n$ of the given state $x$ that have hamming weight in the spike does not surpass $b_\theta = O(\frac{L}{\beta n^\alpha}\ln \theta^{-1})$, then $x\notin \Omega_\theta$. Our approach is to follow the idea of \cite{Crosson2016SimulatedQA} to exploit that adjacent time-slices have highly correlated hamming weights, to present a upper bound on the moment generating function of the spike-time random variable over the spike-less markov chain. This allows us to use a Chernoff bound to bound the odds that the Spike-Time random variable exceeds $b_\theta$, thereby bounding the leakage by an exponential decay.

Finally, we discuss in Section 5 how these new lemmas, under the machinery of \cite{Crosson2016SimulatedQA} can quickly imply a polynomial time algorithm for the entirety of the polynomial region of spike energy gaps, allowing us to conclude with our main result of Theorem \ref{mainresult}.

\section{Simulated Quantum Annealing}
In this section, we present a review of quantum annealing, simulated quantum annealing and the Path Integral Monte Carlo method, and in particular how to construct a (quantum-inspired) classical metropolis-hasting MCMC algorithm to find the ground state of the spike Hamiltonian. The distribution of the markov chain simulates the thermal state $\sigma(s)\propto e^{-\beta H}$ during the quantum annealing procedure, and in subsection 2.2 we bound the distance from sampling from the low temperature thermal state and the actual ground state to define the polynomial inverse temperature $\beta$ that we will simulate the procedure at.

In the ensuing Section 3, we review the key elements of the analysis of \cite{Crosson2016SimulatedQA} on this Markov Chain to bound its convergence time.

\subsection{Quantum Annealing}
Quantum Annealing can be used to minimize an objective $f:\{0 , 1\}^n\rightarrow \mathbb{R}$ by mapping the function onto a diagonal Hamiltonian $H_f=\sum_{x\in \{0, 1\}^n}f(x)|x\rangle\langle x|$ in the computational basis and relaxing the system into the ground state of $H_f$. This relaxation is performed by interpolating $H_f$ with a known, ``easy to prepare" initial Hamiltonian $H_0$, s.t. 
\begin{equation}
    H(s) = (1-s)H_0+s H_f
\end{equation}
over some \textit{knob parameter} $s$ that is slowly tuned during the algorithm from $s=0$ to $1$. It is natural to pick an initial Hamiltonian $H_0 = -\sum_i \sigma_x^i$ corresponding to a uniform transverse field, as one can initialize the system at to the ground of $H_0$ corresponding to the uniform superposition over all computational basis states, 
\begin{equation}
    |\psi_0\rangle = \frac{1}{2^{n/2}}\sum_x |x\rangle
\end{equation}
Following the adiabatic theorem, the final state $|\psi_f\rangle = |\psi(s=1)\rangle$ of the system above after evolving for some time $T$, has high overlap with the ground state of $H_f$ provided $T\geq \text{poly}(n, \delta^{-1})$, where $\delta = \min_s\delta(s)$ is the minimum gap between excited and ground state energies $\delta(s)$ over the entire algorithm. Following the results of Brady and van dam \cite{Brady2016SpectralgapAF}, in the spike regime of $\alpha+2\eta<1$ the spectral gap decays polynomially, and therefore QA converges in polynomial time. However, when $\alpha+2\eta>1$, the gap is exponentially decreasing and convergence takes exponential time.

At any non-zero temperature $\beta<\infty$ we can frame this algorithm in terms of the evolution of an equilibrium thermal state $\sigma(s)\propto e^{-\beta H(s)}$. If the inverse temperature $\beta=\text{poly}(n)$ is sufficiently small and the gap is sufficiently large, the thermal system stays near equilibrium and we can properly account for transitions out of the ground state. It follows that it suffices for the quantum annealing algorithm to sample from $\Pi(z)=\langle z| \sigma(s)|z\rangle$. In the following subsection we analyze the error of this thermal relaxation in the different $\alpha, \eta$ regimes of spikes.

\subsection{Thermal Sampling Error}
Revisiting the thermal sampling error in the different spike regimes is one of our key modifications to \cite{Crosson2016SimulatedQA}'s original proof. In particular, so long as the trace distance $||\sigma(s)-|\psi_0\rangle\langle \psi_0|||_1$ between the thermal state $\sigma(s)\propto e^{-\beta H(s)}$ and the instantaneous g.s. $|\psi_0\rangle\langle \psi_0|$ is sufficiently small then we can approximately sample from a thermal distribution close to the g.s. by sampling from $\sigma(s)$. \cite{Brady2016SpectralgapAF} showed analytical expressions for the energy gap $\Delta$ of the spike Hamiltonian in all regimes of $(\alpha, \eta)$ and \cite{Crosson2016SimulatedQA} used the fact that in the regime of $\alpha+\eta<1/2$ the gap is a constant $\Delta = \Theta(1)$, together with symmetries in the ground state, to show that the error in the thermal relaxation is exponentially decreasing, i.e. 
\begin{equation}
    ||\sigma(s)-|\psi_0\rangle\langle \psi_0|||_1\leq\exp\{-\text{poly}(n)\}
\end{equation}
\noindent for any polynomial inverse temperature. In this section we present very loose bounds on the thermal error, to show the existence of a inverse temperature $\beta$ that allows for exponentially decreasing sampling error in the more general spike regime we are interested in. In particular, we use the following lemma to bound the distance between the distributions

\begin{lemma} \label{thermalerror}
Consider a quantum system on $n$ qubits with Hamiltonian $H$ of energy gap $\Delta$, at some inverse temperature $\beta$. The trace distance between the state in thermal equilibrium $\sigma\propto e^{-\beta H}$ and the ground state $|\psi_0\rangle$ satisfies the following bounds:
\begin{equation}
    \frac{2}{1+e^{\beta \Delta}} \leq ||\sigma-|\psi_0\rangle\langle\psi_0|||_1\leq  2^{n+1}e^{-\beta \Delta}
\end{equation}
\end{lemma}

\noindent \textbf{Proof} The definition of the energy gap tells us $1+e^{-\beta\Delta}\leq \text{Tr}[e^{-\beta H}]\leq 1+2^n e^{-\beta \Delta}$. Note that in the basis of $H$ eigenstates $\sigma-|\psi_0\rangle\langle\psi_0|$ is diagonal, and therefore its 1-norm is the sum of the absolute values of the entries, s.t. $||\sigma-|\psi_0\rangle\langle\psi_0|||_1 = 2|\frac{\text{Tr}[e^{-\beta H}]-1}{\text{Tr}[e^{-\beta H}]}|$. \\

In the spike regime of $\alpha+2\eta\leq 1$ and $\alpha+\eta > 1/2$, the gap is polynomially decreasing \cite{Brady2016SpectralgapAF} $\Delta = \Theta(n^{\frac{1}{2}-\alpha-\eta})$. This implies that the distance 
\begin{equation} \label{polythermal}
    ||\sigma(s)-|\psi_0\rangle\langle \psi_0|||_1\leq 2^n\times e^{-\beta \Delta} \leq \exp\{n\ln 2 - \beta n^{\frac{1}{2}-\alpha-\eta}\} \leq e^{-\Omega(n)}
\end{equation}
is exponentially decreasing for any $\beta = \Omega(n^{1/2+\alpha+\eta})$, and thus the low temperature thermal equilibrium QA converges in polynomial time and allows for approximately sampling from the g.s. of $H(s)$. We note that there is a straightforward lower bound on the sampling error in the exponential regime, where $\alpha+2\eta>1$, showing the impossibility of thermally sampling close to the ground state. Once again as presented in \cite{Brady2016SpectralgapAF}, the gap $\Delta = \text{poly}(n)\exp\{-n^{(\alpha+2\eta-1)/2}\}$ is exponentially decreasing in this regime, and therefore the sampling error becomes too large:
\begin{equation}
    ||\sigma(s)-|\psi_0\rangle\langle \psi_0|||_1\geq  \frac{2}{1+e^{\beta \Delta}}  =  \frac{2}{1+e^{e^{-\text{poly}(n)}}} = 1-o(1)
\end{equation}
which is effectively a constant for any polynomially large inverse temperature.

\subsection{Time Slices and Trotterization}
Our goal now is to construct a classical algorithm that can sample from a distribution close to the thermal equilibrium distribution $\Pi(z)=\langle z| \sigma(s)|z\rangle$ described above. We begin by writing the quantum partition function as an imaginary-time path integral over an exponential number of trajectory basis states, 
\begin{equation}
    Z(s) = \text{tr} e^{-\beta H} = \text{tr}\prod_i^L e^{-\beta H/L} = \sum_{x_1, \cdots x_L\in \{0, 1\}^n} \prod_i\langle x_i| e^{-\beta H/L}|x_{i+1}\rangle
\end{equation}

where we introduce $L=\text{poly}(n, \beta)$ ``time-slices" $x_i\in \{0, 1\}^n$. In this manner, the exponent of each product above $\beta H/L$ can be made sufficiently small under the choice of $L$, and we can trotterize the non-commuting terms under in the definition of $H$ to further analyze the products. Following the discussion in \cite{Crosson2016SimulatedQA} and \cite{Crosson2018RapidMO}, we can establish a $\delta$-multiplicative approximation $\in (1\pm \delta) Z(s)$ to the partition function if we pick $L=\Theta((\beta ||H||)^{3/2}/\delta^{1/2})$, and thereby 
\begin{gather}
    Z(s)= \sum_{x_1, \cdots x_L} \prod_i\langle x_i| e^{-\beta H/L}|x_{i+1}\rangle\approx \sum_{x_1, \cdots x_L} \prod_i\langle x_i| e^{-\beta sH_f/L}e^{-\beta (1-s)H_0/L}|x_{i+1}\rangle = \\
    = \sum_{x_1, \cdots x_L}  e^{-\beta s \sum_i f(x_i)/L }\prod_i\langle x_i| e^{-\beta (1-s)H_0/L}|x_{i+1}\rangle = \sum_{x_1, \cdots x_L}  e^{-\beta s \sum_i f(x_i)/L }\prod_i^L\prod_j^n\langle x_{j,i}| e^{\frac{\beta (1-s)}{L}\sigma^x_j}|x_{j, i+1}\rangle
\end{gather}
where $|x_{j, i}\rangle$ is the $j$th spin of the $i$th time-slice. If we pick $\delta = O(1/n)\Rightarrow L =\Theta(n^2\beta^{3/2})$, we can subsequently ignore the multiplicative approximation to the partition function as it won't affect convergence time or correctness. Finally, note that exp$(\sigma^x)$ can be expanded s.t. up to constant factors
\begin{equation}
    Z(s) = \sum_{x_1, \cdots x_L}  e^{-\beta s \sum_i f(x_i)/L }\prod_i^L\prod_j^n(\delta_{x_{j, i}, x_{j, i+1}}+ (1-\delta_{x_{j, i}, x_{j, i+1}} )\tanh(\omega))
\end{equation}
where $\omega = \beta(1-s)/L$ and $\delta_{a, b}$ is the Kronecker delta, set to 1 if $a=b$. We can view $Z$ as the normalizing constant of a distribution $\pi(x)$ over the state space $x=(x_1\cdots x_L)\in \Omega \equiv \{0, 1\}^{nL}$, where
\begin{equation} \label{stationary}
    \pi(x_1\cdots x_L) = \frac{1}{Z}e^{-\frac{\beta s}{L}\sum_i^L f(x_i)} \times \prod_{i,l=1}^{L, n} (\delta_{x_{j, i}, x_{j, i+1}}+ (1-\delta_{x_{j, i}, x_{j, i+1}} )\tanh(\omega))
\end{equation}
s.t. the probability of a given state $x$ depends on the function values $f(x_i)$ of its time-slices, and since $\tanh \omega < 1$,  $\pi$ also decays with the number of bit-flips $x_{j, i}\neq x_{j, i+1}$ in the \textit{worldline} $(x_{j, 1}, x_{j, 2}\cdots x_{j, L})$ of the $j$th spin. The key point here is that $\Pi(z)$, the thermal distribution we wish to sample from, is approximately the marginal of the distribution of $\pi(x)$ over the additional time-slices, that is, 
\begin{equation}
    \Pi(z) = \sum_{x_2\cdots x_L}\pi(x_1=z, x_2, \cdots x_L)
\end{equation}
s.t. it suffices to sample from $\pi$ to sample $\Pi$. In the following section, we show how to use this interpretation and the Metropolis-Hastings algorithm to construct a Monte Carlo algorithm to sample from $\pi$. 

\subsection{The Single-Site Metropolis Updates Markov Chain}
We construct the \textit{single-site metropolis updates} markov chain, of stationary distribution $\pi$ above, as follows. For a given value $s<1$ of the adiabatic parameter, we run the markov chain monte carlo method on the state space $\Omega = \{0, 1\}^{nL}$ with the following transition kernel $\forall x, x'\in \Omega$

\begin{equation}
    P_M(x, x') = \begin{cases}
        \frac{1}{2nL}\min\big\{1, \frac{\pi(x')}{\pi(x)}\big\} & \text{if $x, x'$ differ in exactly 1 bit} \\
        0 & \text{if $x, x'$ differ in more than 1 bit} \\
        1-\sum_{y\neq x}P_M(x, y) & \text{if $x=x'$}
    \end{cases}
\end{equation}

where we can efficiently generate these transitions through a coin flipping protocol: first flip a unbiased coin, and if heads, then pick one of the $nL$ bits of $x\in \Omega$ uniformly at random, and flip it to define the next state $x'$. Finally, we flip a biased coin with probability $\min\big\{1, \frac{\pi(x')}{\pi(x)}\big\}$ to decide whether or not to transition. Note that one can implement all these sampling steps in $O(\log n)$ time. We then discretize over the values of the adiabatic parameter $s$, and generate samples from $\pi$ for all these values of $s$. Overall, the runtime of this algorithm is simply the convergence time of the underlying MCMC algorithm to stationarity over every value of $s$.

\section{The Paths Comparison Method and Leaky Chains}
In this section we review without proof the analysis approach of \cite{Crosson2016SimulatedQA} in bounding the convergence time of the Single-Site Metropolis Updates Markov Chain of the previous section. The outline of the proof follows from a novel application of a path comparison method, by comparing the markov chains $(\pi, P)$ corresponding to the spike hamiltonian and the markov chain $(\tilde{\pi}, \tilde{P})$ corresponding to the spike-less hamiltonian. The key idea is that the spike-less system is straightforward to analyze, however, its comparison to the spike-hamiltonian is quite non-trivial, due to an exponential number of states $x\in \Omega = \{0, 1\}^{nL}$ that are exponentially more likely in the spike-less system, i.e., $\tilde{\pi}(x)\gg \pi(x)$. This is addressed by partitioning the state space $\Omega$ based on the ratio $\tilde{\pi}(x)/\pi(x)$ into a set of good states $\Omega_G$ and bad states $\Omega_B$, and considering paths that are routed strictly through $\Omega_G$ to relate the congestion of the two chains. Finally, a coupling argument with a leaky markov chain restricted to $\Omega_G$ implies a bound on the distance between the distributions of the spike chain $P$ and the substoquastic matrix $P_G = P\mathbb{I}_{x, y\in \Omega_G}$ corresponding to the restriction. We formalize this statement and discussion in the following series of Theorems.

\begin{theorem} [Most-Paths Comparison Method \cite{Crosson2016SimulatedQA}] \label{mostpaths} Let $(\pi, P)$ and  $(\tilde{\pi}, \tilde{P})$ be reversible Markov chains with the same state space graph $(\Omega, E)$. Let $a = \max_{x\in \Omega} \pi(x)/\tilde{\pi}(x)$ and define $\Omega_\theta \equiv \{x\in \Omega: \pi(x)\leq \theta\tilde{\pi}(x)\}$. If there is a set of canonical paths for $(\tilde{\pi}, \tilde{P})$ achieving congestion $\tilde{\rho}$ and satisfying $3a^2\tilde{\rho}\tilde{\pi}(\Omega_\theta) < 1$, then there is a partition $\Omega_G\cup\Omega_B = \Omega$ with $\pi(\Omega_G)\geq 1-3a^2\tilde{\rho}\theta\pi(\Omega_\theta)$, and a canonical flow for $(\pi, P)$ that connects every $x, y \in \Omega_G$ with paths contained in $\Omega_G$ for which the congestion $\rho$ of any edge in $\Omega_G$ is
\begin{equation}
    \rho\leq 16  \max_{x, y\in \Omega_G}\bigg[\frac{\tilde{P}(x, y)}{P(x, y)}\bigg]a^2 \tilde{\rho}\theta^{-1} =  O(\theta^{-1} \tilde{\rho})
\end{equation}

\end{theorem}

\noindent moreover, \cite{Crosson2016SimulatedQA} upper bounded the congestion of the spike-less markov chain under single-site updates $\tilde{\rho} = O(n^5L^5\beta^{-3}) = O(n^{15}\beta^{9/2})$ under the value of $L=\Theta(n^2\beta^{3/2})$ as established in bounding the error due to the trotterization in Section 2.

The key idea behind this statement is that one can find a subset of the state space $\Omega_G$ that is well connected in both the spike and spike-less chains, and upper bound the congestion in the spike chain restricted to $\Omega_G$ by a linear function of the ``easy to compute" spike-less chain congestion. Effectively this implies that both Markov Chains mix fast over the ``good subset", however doesn't yet address the transitions to the  bad subset. \cite{Crosson2016SimulatedQA} then used an elegant coupling argument to compare the following three chains: the \textit{restricted} chain induced by the restriction of the spike chain over $\Omega_G$, a \textit{replacement} chain where any transition out of $\Omega_G$ is replaced back into $\Omega_G$ following the stationary distribution $\pi$, and an \textit{absorbing} chain where any transition out of $\Omega_G$ is to an additional absorbing vertex. The theorem below formalizes this result on a bound on the convergence of the spike chain in terms of the ``leakage" probability $\pi(\Omega_B)$:

\begin{theorem}[Leaky Markov Chains \cite{Crosson2016SimulatedQA}]\label{coupling} Let $(\pi, \Omega, P)$ be a reversible markov chain and suppose $\Omega_G\cup \Omega_B=\Omega$ is a partition. Let $P_G$ be the substoquastic transition matrix $P_G(x, y)=P(x, y)\mathbb{I}_{x\in \Omega_G}\mathbb{I}_{y\in \Omega_G}$. Suppose there is a set of canonical paths connecting every pair of points $x,y\in \Omega_G$, and the congestion of the walk on this set of paths is $\rho$. If $\mu$ is a warm start with $\mu(x)\leq M\pi(x)$ for all $x\in \Omega_G$, then the distribution obtained by starting from $\mu$ and applying $t$ steps of the random walk satisfies
\begin{equation}
||\mu P^t_G-\pi||_1\leq Mt\pi(\Omega_B) + \pi_{min}^{-1}e^{-t/\rho}
\end{equation}
\end{theorem}

And thereby the distance between the restricted walk and the stationary distribution over the set of good states is upper bounded by a sum of two terms: an exponential decay with rate $\rho$ due to the fast mixing over $\Omega_G$, and a constant rate leakage of $M\pi(\Omega_B)\leq O(\tilde{\rho}\tilde{\pi}(\Omega_\theta))$. We defer the discussion on the $M$-warm starts to Section 5.

This reduces the rest of the analysis to showing the existence of a state space partition parametrized in terms of some $\theta = O(1)$ that allows for an intermediary grace period of quasi-stationary mixing, s.t. the leakage probability in the spikeless chain $\tilde{\pi}(\Omega_\theta)$ is exponentially decaying as $e^{-\Omega(n)}$, while the mixing time defined by $\rho = O(\theta^{-1}\tilde{\rho})$ remains polynomial. Then, it would follow that we can indeed sample from a distribution close to the stationary distribution.  We address this issue in the following sections by revisiting the bound on $\tilde{\pi}(\Omega_\theta)$ in \cite{Crosson2016SimulatedQA}'s original work under tighter concentration bounds.

\section{The Leakage Probability}

The sections before reduce the analysis to defining a 'bad subset' of states $\Omega_\theta\subset \Omega\equiv \{0,1\}^{nL}$ that are \textit{relatively unlikely} to be reached under the spike-chain, i.e. picking a threshold $\theta$ s.t.

\begin{equation}
    \Omega_\theta = \{ x\in \Omega \textrm{ s.t. } \frac{\pi(x)}{\tilde{\pi}(x)}< \theta \}
\end{equation}

\noindent where every $x\notin \Omega_\theta$ obeys

\begin{equation}
\frac{\pi(x)}{\tilde{\pi}(x)} \geq  \frac{\tilde{Z}}{Z}\exp\bigg\{-\frac{\beta n^\alpha}{L}\max_{x\notin \Omega_\theta} \textrm{ST}(x)\bigg\} \geq \exp\bigg\{-\frac{\beta n^\alpha}{L}\max_{x\notin \Omega_\theta} \textrm{ST}(x)\bigg\} \geq \theta
\end{equation}

 As mentioned in Section 3, if such a $\Omega_\theta$ can be picked so that the \textit{leakage} $\tilde{\pi}(\Omega_\theta)$ is exponentially small while $\theta$ is at most a constant, then we can sample from a distribution close to the ground state during a quasi-stationary mixing period. Harrow and Crosson \cite{Crosson2016SimulatedQA} do so in the spike regime where the gap is constant by inspecting the \textit{spike time}, defined as follows
\begin{equation}
    ST(x) = \sum_{i=1}^L \mathbb{I}_{R}(x_i)
\end{equation}
where $\mathbb{I}_R$ is the indicator random variable set to 1 if $x_i\in R$, i.e., if the hamming weight of the \textit{i}th worldline is in the spike. The probability of leakage $\tilde{\pi}(\Omega_\theta)$ is then bounded via a moment bound through Markov's inequality 

\begin{equation}
    \tilde{\pi}(\Omega_\theta) = \mathbb{P}[x\in \Omega_\theta]_{\tilde{\pi}} = \mathbb{P}\big[ST(x) \geq b_\theta\big]_{\tilde{\pi}}\leq \frac{\mathbb{E}[ST^m]_{\tilde{\pi}}}{b_\theta^m}
\end{equation}
with $b_\theta = \frac{L}{\beta n^\alpha}\ln \frac{1}{\theta}$. The authors then continue by upper bounding the \textit{m}th moment, however, using an analysis that is only tight for low moments $m=\Theta(1)$. Our strategy will be to apply a tighter concentration bound to the leakage probability, the Chernoff Bound, by computing an upper bound on the moment generating function.
\begin{equation}
    \tilde{\pi}(\Omega_\theta) =\mathbb{P}\big[ST(x) \geq b_\theta\big]_{\tilde{\pi}}= \mathbb{P}\big[e^{\lambda ST(x)} \geq e^{\lambda b_\theta}\big]_{\tilde{\pi}}\leq \frac{\mathbb{E}[e^{\lambda ST}]_{\tilde{\pi}}}{e^{\lambda b_\theta}} 
\end{equation}

\noindent For any positive parameter $\lambda$. In the following subsection \ref{largemom}, we extend the analysis of \cite{Crosson2016SimulatedQA} to larger moments $m$ and use a series expansion to bound the MGF. The main result of this section is the following lemma over the leakage probability

\begin{lemma}[The Leakage Probability]\label{leakageprob}
When $\alpha+2\eta \leq 1$ there exists a threshold $\theta = \Theta(1)$ such that the leakage probability in the spike-less chain is negligible, that is, $\tilde{\pi}(\Omega_\theta)\leq e^{-\Omega(n)}$.
\end{lemma}

In particular, an exponentially small leakage probability means that after a polynomial quasi-stationary mixing time, we can sample from the ground state $|\psi(s)\rangle$ within polynomial error through Theorem \ref{coupling}. In the ensuing Section \ref{effconv}, we follow the discussion in \cite{Crosson2016SimulatedQA} and present the missing technical details before concluding with our main Theorem.

\subsection{A Large Moment Bound}
\label{largemom}
The moment bound obtained by Harrow and Crosson \cite{Crosson2016SimulatedQA} in Lemma \ref{lowmoments} is only tight for low moments $m$. We can obtain an accurate expansion of the moment generating function if we explicitly compute the moments $\mathbb{E}[ST^m]$ in the limit of $m=\Omega(L)$. To extend the analysis to $m>L$, we reduce the upper bound to linear combinations of the smaller moments, and use \cite{Crosson2016SimulatedQA}'s result re-stated here in Lemma \ref{lowmoments}. For concreteness, we rehash their proof of the upper bound on the $m$th moment in the Appendix.
\begin{lemma}
[The Spike-Time Moments \cite{Crosson2016SimulatedQA}]\label{lowmoments}
 $\mathbb{E}[ST^m]_{\tilde{\pi}} \leq (Ln^{\eta-\frac{1}{2}})^m(1+o(1))$
\end{lemma}

Formally, what we want to prove are the following tighter upper bounds on the higher order moments: 

\begin{lemma}[Large Spike-Time Moments]
\label{largemomlemma}
 $\mathbb{E}[ST^m]  \leq 
\begin{cases}
   \big(2Ln^{\eta-\frac{1}{2}}\big)^m & L>m \\
  m^{L}L^{L}n^{L(\eta-\frac{1}{2})} & m \geq 2L \\
  (2eL)^{L}n^{L(\eta-\frac{1}{2})} & m\in [L, 2L-1]
 \end{cases}$\\
\end{lemma}

We do so as follows. The key idea here is that the $m$-point correlation functions now are composed of products of indicator random variables at multiple repeated times, and thus we should individually count the number of distinct worldlines $l$ in the product, sum over all possibilities $a_1, a_2\cdots a_l\in [L]$ of those worldlines, and sum over the distinct distribution of repeated counts $c_{a_1}\cdots c_{a_l}$ totalling $m$:
\begin{equation}
    \mathbb{E}[ST^m] = \sum_{t_1, t_2\cdots t_m}^L\mathbb{E}[\prod_i^m \mathbb{I}_R(z_{t_i})] = \sum_{l=1}^{\min(m, L)}\sum_{a_1<a_2\cdots a_l}^L\sum_{\substack{c_{a_1}\cdots c_{a_l}\geq 1\\ \sum_i c_{a_i}=m}}\mathbb{E}[\prod_i^l \mathbb{I}_R(z_{a_i})]
\end{equation}
We exchange the inner-most summation as the number of integer solutions to the equation $\sum_{i=1}^l c_{a_i}=m$ with each $c_{a_i}\geq 1$ is $\binom{m-1}{l-1}$.
\begin{equation}
    \sum_{a_1<a_2\cdots a_l}^L\sum_{\substack{c_{a_1}\cdots c_{a_l}\geq 1\\ \sum_i c_{a_i}=m}}\mathbb{E}[\prod_i^l \mathbb{I}_R(z_{a_i})] =  \binom{m-1}{l-1}\sum_{a_1<a_2\cdots a_l}^L\mathbb{E}[\prod_i^l \mathbb{I}_R(z_{a_i})]\leq \binom{m-1}{l-1}\mathbb{E}[ST^l]
\end{equation}
where the $l$-th moment above appears as a relaxation over the constraint on the overlapping $a_i$'s. Now, we can just use the expression in Lemma \ref{lowmoments} \cite{Crosson2016SimulatedQA} for low moments $l < L$. 
\begin{gather}
    \mathbb{E}[ST^m] = \sum_{l=1}^{\min(m, L)} \binom{m-1}{l-1}\mathbb{E}[ST^l] \leq \sum_{l=1}^{\min(m, L)} \binom{m-1}{l-1} L^l n^{l(\eta-\frac{1}{2})} \leq \\ \leq 
\begin{cases}
   \big(2Ln^{\eta-\frac{1}{2}}\big)^m & L>m \\
  m^{L}L^{L}n^{L(\eta-\frac{1}{2})} & m \geq 2L \\
  (2eL)^{L}n^{L(\eta-\frac{1}{2})} & m\in [L, 2L-1]
 \end{cases}
\end{gather}
where we consider apart the case $[L, 2L-1]$ to address the peak of the binomial term, by upper bounding the binomial by $(em/l)^l$. We note that the dependence on $m$ is consistent with the original bound for all $m< L$, and that the result is monotonically increasing with $m$. This concludes the proof of Lemma \ref{largemomlemma}. We can now follow suit and present an upper bound on the moment generating function via a series expansion. 

\begin{lemma}
[The Spike-Time MGF] $\mathbb{E}[e^{\lambda ST}] \leq \exp\{\Theta(\lambda + L\log \lambda +L\log n)\}$
\end{lemma}

We compute the MGF above by splitting the summation over $m$ into those 3 regimes:
\begin{equation}
    \mathbb{E}[e^{\lambda ST}] \leq \sum_{m=0}^L \frac{\lambda^m}{m!}\big(2Ln^{\eta-\frac{1}{2}}\big)^m+L\times\frac{\lambda^{2L}}{L!} (2eL)^{L}n^{L(\eta-\frac{1}{2})} + L^{L}n^{L(\eta-\frac{1}{2})}\sum_{m\geq 2L} m^{L}\frac{\lambda^m}{m!}
\end{equation}
We address the three terms above individually. The first term ($m<L$),
\begin{equation}
    \sum_{m=0}^L \frac{1}{m!}\big(2\lambda Ln^{\eta-\frac{1}{2}}\big)^m \leq \sum_{m=0}^L \big(2\lambda Ln^{\eta-\frac{1}{2}}\big)^m = \Theta\big((2\lambda Ln^{\eta-\frac{1}{2}})^L\big) = \exp\{L\cdot \Theta(\log \lambda +\log n)\}
\end{equation}
for large enough $\lambda= \Omega(L^{-1}n^{ \frac{1}{2}-\eta})$. The $m\in [L, 2L-1]$ term, 
\begin{equation}
    L\times\frac{\lambda^{2L}}{L!} (2eL)^{L}n^{L(\eta-\frac{1}{2})}\leq \exp\{L\cdot \Theta(\log \lambda +\log n)\}
\end{equation}
And finally the bound over the $m\geq 2L$ term. Note first $m!\geq em^me^{-m}$
\begin{gather}
    L^{L}n^{L(\eta-\frac{1}{2})}\sum_{m\geq 2L} m^{L}\frac{\lambda^m}{m!}\leq \frac{1}{e}L^{L}n^{L(\eta-\frac{1}{2})}\sum_{m\geq 2L} m^{L}\frac{(e\lambda)^{m}}{m^m} =\\=\frac{1}{e}L^{L}n^{L(\eta-\frac{1}{2})}(e\lambda)^{2L}\sum_{m'\geq 0} \frac{(e\lambda)^{m'}}{(m'+2L)^{m'}} \leq \frac{1}{e}L^{L}n^{L(\eta-\frac{1}{2})}(e\lambda)^{2L}e^{e\lambda }\leq \\
    \leq \exp\{\Theta(\lambda + L\log \lambda +L\log n)\}
\end{gather}
and evidently this term is the largest, concluding the proof of the lemma. \\

We can now apply it to the Chernoff bound mentioned in the introduction to this section, and prove Lemma \ref{leakageprob} on the Leakage Probability. If we require $\theta^{-1} = \Theta(1)$
\begin{equation}
    \mathbb{P}\bigg[ST(x)>b_\theta = O(\frac{L}{\beta n^\alpha}\log \theta^{-1})\bigg]\leq \exp\{\Theta(\lambda + L(\log \lambda+\log n) - \frac{\lambda L}{n^\alpha\beta})\}= e^{-\Omega(n)}
\end{equation}
so long as $\lambda = \Omega(n^\alpha\beta\log n)$. It follows that for $\alpha+2\eta < 1$ and $\theta = \Theta(1)$ we have $\tilde{\pi}(\Omega_\theta) = e^{-\Omega(n)}$, and therefore the rate of probability leakage to the subset of bad states $\Omega_B$ is exponentially small. This concludes the proof of Lemma \ref{leakageprob}. \\

In addition to Theorem $\ref{mixing}$, this result implies a period of quasi-stationary mixing times where the marginal distributions $\Pi(z)$ of the single-site random walk within the good subset $\Omega_G$ is exponentially close to the stationary distribution, thereby allowing us to efficiently sample from the ground state. We conclude this paper by presenting the concrete numerical runtime of our algorithm, following the dicussion in \cite{Crosson2016SimulatedQA}.

\section{Efficient Convergence} \label{effconv}
In this section we follow the discussion in \cite{Crosson2016SimulatedQA} and present the missing technical details and numerics behind the runtime of the SQA algorithm presented in Section 2. As mentioned in Section 2.4 and 3, we have yet to argue how to discretize the adiabatic schedule, how to initialize the warm starts, and finally to present precise bounds on the (quasi-) stationary mixing time. 

Following the presentation in \cite{Crosson2016SimulatedQA}, we discretize the adiabatic schedule from $s_{min}=0, s_1, s_2\cdots$ to $s_{max}=1-\delta s$ with a step size of $\delta s = \tilde{O}((n\beta)^{-1})$. This enforces that the ground state at $s=s_{max}$ is sufficiently close to that of $s=1$
\begin{equation}
    ||\psi_0(s=1)\rangle-|\psi_0(s=s_{max})\rangle |\leq \text{poly}(n)^{-1}
\end{equation}
and moreover guarantees the warm start condition with $M=2$: that is, for each value $s=s_i$, we run the Markov Chain until the quasi-stationary regime - s.t. the starting distribution of the next iteration $\mu\approx \pi_i$ is $\leq 2\pi_{i+1}$ $\forall x\in \Omega_G$.

Now, it only remains to show the duration of the quasi-stationary regime of the single-site Markov Chain of Section 2. We require the variational distance of the random walk distribution $\mu P^t$ to the stationary distribution $\pi$ to be exponentially small $e^{-\Omega(n)}$. This is possible so long as we pick a time $t$ s.t. the expression
\begin{equation}
    ||\mu P^t_G-\pi||_1 \leq O(te^{-\Omega(n)}+\pi_{min}^{-1}e^{-t/\rho})
\end{equation}
is $e^{-\Omega(n)}$. This inequality defines the quasi-stationary mixing, and it suffices to pick $t = O(n\times\rho \log\pi_{min}^{-1})=O(n\tilde{\rho} \log \pi_{min}^{-1}) = O(n^{16}\beta^{9/2}\log \pi_{min}^{-1})$, where Theorem \ref{mostpaths} relates the congestions $\rho = O(\theta \tilde{\rho}) = O(\tilde{\rho})$ of the spike and spike-less markov chains, and the polynomial upper bound $\tilde{\rho}=O(n^{15}\beta^{9/2})$ is discussed in section 3.

\cite{Crosson2016SimulatedQA} then point out that under the stationary distribution of equation \ref{stationary}, in each worldline the number of adjacent bits that differ follows a binomial distribution of mean $\Theta(\beta)$, and concentrates around the mean. In this manner, conditioning on the exponentially low probability event that any worldline differs in more than $\tilde{O}(\beta)$ adjacent bit flips not happening, the minimum value of $\pi$ can be lower bounded to $O(\exp\{-\beta n\log n\})$. It follows $\log \pi_{min}^{-1} = \tilde{O}(n\beta)$, and therefore our quasi-stationary mixing time becomes $\tilde{O}(n^{16}\beta^{11/2})$. 

To conclude, we note that in Section 2 we argued how to implement each single-site MCMC step in $O(\log n)$ time, and in the beginning of this section we discussed the bound of $\tilde{O}(n\beta)$ on the length of the adiabatic schedule. Overall, the runtime becomes $\Tilde{O}(n^{17}\beta^{13/2}) = O(n^{37})$ once we pick a small enough temperature $\beta = \Theta(n^{1/2+\alpha+\eta}) = O(n^{3/2})$. Formally,

\begin{theorem}
Simulated Quantum Annealing based off of the path-integral Monte Carlo method efficiently samples the output distribution of QA for the spike cost function when $\alpha+2\eta\leq 1$. The running time using single-site spin flips is $O(n^{37})$.
\end{theorem}

\section{Acknowledgements}
The author would like to thank Aram Harrow, Isaac Chuang, Yongshan Ding, and John Napp for their guidance and suggestions over the course of this research. \\
This work was partially supported with
funding from NSF grant PHY-1818914.

\newpage
\printbibliography

\newpage
\begin{appendices}
\section{The $m$ Time Slice Correlations \cite{Crosson2016SimulatedQA}}
\label{lowmoment}
Harrow and Crosson upper bounded the $m$th moment of the spike time, in order to apply a concentration bound to the leakage probability. Their upper bound was limited to low moments $m=\Theta(1)$, and is key to our extension to larger moments $m=\Omega(L)$ (around the system size). We formalize their result in the following lemma, and sketch their proof here. 

\begin{lemma}
[The Spike-Time Moments \cite{Crosson2016SimulatedQA}]
 $\mathbb{E}[ST^m]_{\tilde{\pi}} \leq (Ln^{\eta-\frac{1}{2}})^m(1+o(1))$
\end{lemma}

The mth moment can be written as a sum of the m-point correlation functions on the worldline. This can be approximated to subleading error $O(L^{-1})$ to the expectation over the ground state of the spike-less chain of a projector operator on the spike. In particular, define the basis $\{|k\rangle:k\in \{0\cdots n\}\}$ for the symmetric subspace, labelled by hamming weight, and define the operator $S = \sum_{k\in R} |k\rangle \langle k|$ such that:
\begin{equation}
    \mathbb{E}[ST^m]_{\tilde{\pi}} = \sum_{t_1, t_2\cdots t_m}^L \mathbb{E}[\mathbb{I}_R(z_{t_1})\mathbb{I}_R(z_{t_2})\cdots \mathbb{I}_R(z_{t_m})] = \sum_{t_1, t_2\cdots t_m}^L \langle e^{-\tau_1 H}Se^{-(\tau_2-\tau_1) H}S\cdots \rangle
\end{equation}
with $\tau_i=\frac{\beta t_i}{L}$. This is then expanded under the basis of symmetric energy eigenstates of the spike-less system $\{|\tilde{\psi}_k\rangle\}$ obtaining 
\begin{equation}
    \mathbb{E}[ST^m]_{\tilde{\pi}} = \sum_{t_1\cdots t_m , k_1\cdots k_m}\bigg(\prod_i e^{-(\tau_{i+1}-\tau_i)\Delta k_i} \langle \tilde{\psi}_{k_{i+1}}|S|\tilde{\psi}_{k_i}\rangle \bigg)
\end{equation}
where $\Delta = O(1)$ is the gap of the spike-less Hamiltonian. The authors then note that the spike time can only be large when the peak of the g.s. is \textit{near} the spike $R$. In this regime, the spike-less eigenstates satisfy the upper bound $\langle \tilde{\psi}_{k}|w\rangle \leq |\langle \tilde{\psi}_0|w\rangle | = O(n^{-\frac{1}{4}})\forall k\in [n]$ and $w\in R$.  This allows for an upper bound on the equation above 
\begin{equation}
    \mathbb{E}[ST^m]_{\tilde{\pi}}\leq n^{m(\eta-\frac{1}{2})}\times \sum_{t_1\cdots t_m , k_1\cdots k_m}\bigg(\prod_i e^{-(\tau_{i+1}-\tau_i)\Delta k_i} \bigg)
\end{equation}
where the $t_i\in [L]$ and are ordered, and the $k_i\in [n]$ and are not ordered. Harrow and Crosson then procede by redefining $t_{i+1}-t_i=g_i$ in the summation above and relaxing the constraint $\sum_i g_i = L$ to $g_i\in [L]\forall i$. The leading order dependence can then be extracted by dividing into cases on the number of $k_i$'s that are 0, i.e.
\begin{gather}
    \sum_{t_1\cdots t_m , k_1\cdots k_m}\bigg(\prod_i e^{-(\tau_{i+1}-\tau_i)\Delta k_i} \bigg) = \sum_{l=1}^m \binom{m}{l}\frac{L^{m-l}}{(m-l)!}\times \sum_{g_{a_1}\cdots g_{a_l},k_{a_1}\cdots k_{a_l}}\bigg(\prod_i e^{-g_i\beta\Delta k_i/L} \bigg) \\
    \leq L^m \sum_{l=1}^m \binom{m}{l}(\beta\Delta)^{-l} \sum_{k_{a_1}\cdots k_{a_l}}\prod_i\frac{1}{k_{a_i}} \leq L^m \sum_{l=1}^m \bigg(\frac{1+\log n}{\beta\Delta m}\bigg)^l
\end{gather}
where in the first equality we perform the summation over the $g$'s corresponding to zero $k$'s, the first inequality arises from the summation over the g's corresponding to nonzero k's, and finally the last inequality arises from l products of harmonic numbers. Overall, the $m$-th moment bound encountered by the authors was

\begin{equation}
     \mathbb{E}[ST^m]_{\tilde{\pi}}\leq L^m n^{m(\eta-\frac{1}{2})}\sum_{l=1}^m \bigg(\frac{1+\log n}{\beta\Delta m}\bigg)^l
\end{equation}
we note that the geometric series $\sum_{l=1}^m \bigg(\frac{1+\log n}{\beta\Delta m}\bigg)^l\leq \frac{1}{1-\frac{1+\log n}{\beta \Delta m}}$ since $\beta = \textrm{poly}(n)$, as mentioned in Section 2.1. Note that for $m=1$, we obtain $\mathbb{E}[ST(x)]_{\tilde{\pi}} \leq Ln^{\eta-\frac{1}{2}}$, which is \textit{tight} in the sense that $\mathbb{E}[ST(x)]_{\tilde{\pi}}=\Theta(Ln^{\eta-\frac{1}{2}})$ for certain pick of $s$ during the execution of the algorithm. This concludes the proof.

\end{appendices}

\end{document}